\documentstyle[preprint,revtex,eqsecnum]{aps}
\tightenlines
\begin{document}
%\draft

\begin{flushright}
\vspace*{-2cm}
TUTP-93-1  \\ February 1993
\vspace*{2cm}
\end{flushright}

\begin{title}
SEMICLASSICAL GRAVITY THEORY\\
AND QUANTUM FLUCTUATIONS
\end{title}
\author{Chung-I Kuo and L. H. Ford}
\begin{instit}
Institute of Cosmology\\
Department of Physics and Astronomy\\
Tufts University\\
Medford, Massachusetts 02155
\end{instit}

\begin{abstract}
We discuss the limits of validity of the semiclassical theory of
gravity in which a classical metric is coupled to the expectation
value of the stress tensor. It is argued that this theory is a
good approximation only when the fluctuations in the stress tensor
are small. We calculate a dimensionless measure of these fluctuations
for a scalar field on a flat background in
particular cases, including squeezed states
and the Casimir vacuum state. It is found that the fluctuations
are small for states which are close to a coherent state, which
describes classical behavior, but tend to be large otherwise. We
find in all cases studied that the energy density fluctuations
are large whenever the local energy density is negative. This is
taken to mean that the gravitational field of a system with negative
energy density, such as the Casimir vacuum, is not described by a fixed
classical metric but is undergoing large metric fluctuations. We
propose an operational scheme by which one can describe a fluctuating
gravitational field in terms of the statistical behavior of test particles.
For this purpose we obtain an equation of the form of the Langevin
equation used to describe Brownian motion.
\end{abstract}

\section{Introduction}

A natural proposal to describe the gravitational field of a quantum
system is the semiclassical theory in which the expectation value of the
stress tensor is the source. The semiclassical Einstein equation is
\begin{equation}
G_{\mu\nu}\equiv
R_{\mu\nu}-{1\over 2}g_{\mu\nu}R
=8 \pi G_N\langle T_{\mu\nu} \rangle. \label{eq:semi}
\end{equation}
This theory is almost certain to fail at the Planck scale, where the
quantum nature of gravity becomes important. However, it can also fail
far away from the Planck scale if the fluctuations in the stress
tensor become important. This was discussed some time ago by one of
us \cite{Ford82} and illustrated by the problem of graviton
emission by a box of particles in a general quantum state.
It was found that the semiclassical theory gives reliable results
when the fluctuations in the stress tensor are not too large,
that is, when
\begin{equation}
\langle T_{\alpha \beta}(x) \,T_{\mu \nu}(y) \rangle \approx
\langle T_{\alpha \beta}(x) \rangle
\langle T_{\mu \nu}(y) \rangle.
\end{equation}
However, for quantum states in which the energy density fluctuations
are large, the semiclassical theory based upon Eq.(\ref{eq:semi})
cannot be trusted.

In the present paper, we will further explore the issue of the limits
of validity of the semiclassical theory. Particular attention will be
paid to quantum states in which the expectation value of the local
energy density can be negative. In Sec.~\ref{sec:negative} some
aspects of such states will be reviewed, partly for the purpose of
establishing notation and results which will be used later. In
Sec. ~\ref{sec:limits} the range of validity of the semiclassical theory
will be probed in particular flat space examples,
including squeezed states
and the Casimir effect. In Sec.~\ref{sec:test} a proposal will
be made as to how to describe the gravitational field of a system
in which the energy density fluctuations are large. We suggest that
a statistical description of the Brownian-like motion of test
particles in such a gravitational field yields all of the information
which is available about the system. Finally, in Sec.~\ref{sec:last}
the results will be summarized and discussed.

We adopt the convention $ c=\hbar=1 $ and
a spacelike metric $ \eta_{\mu\nu}=diag\lbrack-1,1,1,1\rbrack $
throughout this paper.

\section{Quantum States of Negative Energy Density}
\label{sec:negative}

Although all known forms of classical matter
have non-negative energy density, it is not so
in quantum field theory. A general state can be a
superposition of number eigenstates
and may have a negative expectation value
of energy density in certain spacetime regions
due to coherence effects, thus
violating the weak energy condition
\cite{Epstein65}. If there were no constraints
on the extent of the  violation of the weak energy
condition, several dramatic and disturbing
effects could arise. These include the
breakdown of the second law of thermodynamics
\cite{Ford78}, of cosmic censorship
\cite{Roman90},  and of causality \
cite{Morris88}. There are, however, two
possible reasons as to why these effects will
not actually arise. The first is the existence
of constraints on the magnitude and the spatial
or temporal extent of the negative energy
\cite{Ford78,Roman90,Ford90}. The second is
that the semiclassical theory of gravity may
not be applicable to systems in which the
energy density is negative. This latter
possibility will be the main topic to be
investigated in this paper.

Let us consider a massless, minimally coupled
scalar field for which the Lagrangian density
is
\begin{equation}
{\cal L}={1\over 2}\eta_{\mu\nu}
(\partial^\mu\phi)(\partial^\nu\phi).
\end{equation}
The stress tensor is
\begin{equation}
T_{\mu\nu}=\Pi(\partial_0\phi)-
\eta_{\mu\nu}{\cal L}=
(\partial_\mu\phi)(\partial_\nu\phi)-
{1\over 2}\eta_{\mu\nu}
(\partial_\sigma\phi)(\partial^\sigma\phi),
\end{equation}
where $ \Pi = \partial_{0}\phi $ is the
conjugate momentum of $\phi$.
Variation with respect to $\phi$ gives the
dynamical equation
\begin{equation}
(-{\partial^2\over\partial t^2}+\nabla^2)
\phi(x) \equiv \partial_\mu\partial^\mu
\phi(x)=0.
\end{equation}
The quantum field operator $\phi$ may
be expanded in mode functions as
\begin{equation}
\phi=\sum_{\bf k}(a_{\bf k}f_{\bf k}
+a_{\bf k}^{\dagger}f_{\bf k}^{\ast}),
\end{equation}
where
\begin{equation}
\lbrack a_{\bf k},a_{\bf k'}^{\dagger}
\rbrack=\delta_{\bf k \bf k'}, \qquad
\lbrack a_{\bf k},a_{\bf k'}\rbrack
=\lbrack a_{\bf k}^{\dagger},a_{\bf k'}
^{\dagger}\rbrack=0,
\end{equation}
and the mode function is a solution of the
dynamical equation
\begin{equation}
f_{\bf k}=(2L^{3}\omega)^{-1/2}
e^{ik_{\sigma} x^{\sigma}}=
(2L^{3}\omega)^{-1/2}
e^{i{\bf k}\cdot{\bf x}-i\omega t}.
\end{equation}
Here we have assumed periodic
boundary conditions in a 3-dimensional
box of side L. The sum
$ \sum_{\bf k} $ is transformed to
$ {{L^3}\over (2\pi)^{3}}\int d^3 k $
in the continuum limit.
The mode functions are normalized on a
spacelike hypersurface under the scalar
product
\begin{equation}
(\phi_1,\phi_2)\equiv
-i\int d^{n-1}x \{\phi_{1}(x)
\partial_{0}\phi_{2}^{\ast}(x)-
(\partial_{0}\phi_{1}(x))\phi_{2}^{\ast}(x)\},
\end{equation}
so that
\begin{equation}
(f_{\bf k},f_{\bf k'})= \delta_{\bf k \bf k'}.
\end{equation}

We will here consider states where a single mode is excited.
Such a quantum state takes the form
\begin{equation}
 |\Psi\rangle =\sum_{n=0}^{\infty} c_n|n\rangle,
\end{equation}
where $|n\rangle$ is a number eigenstate with $n$ particles in mode ${\bf k}$
and the $c_n$ are any coefficients such that $\sum_{n=0}^{\infty} |c_n|^2 =1$.
The normal ordered expectation value of the
stress tensor is
\begin{eqnarray}
\langle\colon
T_{\alpha\beta}(x)\colon\rangle=&&
\langle\Psi|\colon T_{\alpha\beta}\colon
|\Psi\rangle \nonumber \\
=&&\sum_{n=0}^{\infty}\big(2n|c_n|^2
T_{\alpha\beta}[f_{\bf k},f_{\bf k}^\ast]
+n^{1/2}(n-1)^{1/2}c_n c_{n-2}^{\ast}
T_{\alpha\beta}[f_{\bf k},f_{\bf k}]
\nonumber \\
&&+n^{1/2}(n-1)^{1/2}c_n^{\ast} c_{n-2}
T_{\alpha\beta}[f_{\bf k}^{\ast}, f_{\bf
k}^{\ast}] \big),
\end{eqnarray}
where $T_{\alpha\beta}[g,h]$ is the bilinear
form
\begin{equation}
T_{\mu\nu}[g,h](x)= (\partial_\mu
g)(\partial_\nu h)- {1\over 2}\eta_{\mu\nu}
(\partial_\sigma g)(\partial^\sigma h).
\end{equation}
The different bilinear forms arising from the
mode functions are
\begin{equation}
T_{\alpha\beta}[f_{\bf k},f_{\bf k}]
=-{\cal K}_{\alpha\beta}e^{2i\theta},
\end{equation}
\begin{equation}
T_{\alpha\beta}[f_{\bf k}^{\ast},f_{\bf k}]
=T_{\alpha\beta}[f_{\bf k},f_{\bf k}^{\ast}]
={\cal K}_{\alpha\beta},
\end{equation}
and
\begin{equation}
T_{\alpha\beta}[f_{\bf k}^{\ast},
f_{\bf k}^{\ast}]
=-{\cal K}_{\alpha\beta}e^{-2i\theta},
\end{equation}
where $\theta\equiv k_{\rho}x^{\rho}$ and
${\cal K_{\alpha\beta}}\equiv
(k_{\alpha}k_{\beta}-{1\over 2}
\eta_{\alpha\beta}k_{\rho}k^{\rho})/(2\omega L^{3})$. For the
massless case $k_{\rho}k^{\rho}=0$, so
${\cal K_{\alpha\beta}}=
{k_{\alpha}k_{\beta}\over
(2\omega L^{3})}$.
 Notice that
$\colon T_{\alpha\beta}\colon=
T_{\alpha\beta}-\langle 0|\,T_{\alpha\beta}
\,|0\rangle$; the
normal ordered stress tensor in flat
spacetime is the renormalized result obtained
by subtracting the Minkowski vacuum
expectation value.

The quantum coherence effects which produce
negative energy densities can be easily
illustrated by the state composed of two
particle number eigenstates
\begin{equation}
|\Psi\rangle\equiv
{1\over\sqrt{1+\epsilon^2}
} (|0\rangle + \epsilon|2\rangle), \label{eq:vacplus2}
\end{equation}
where $|0\rangle$ is the vacuum state
satisfying $a|0\rangle=0$,
and $|2\rangle=
{1\over\sqrt 2}(a^{\dagger})^2|0\rangle$
is the two particle state. Here we take $\epsilon$, the relative
amplitude of the two states, to be real for simplicity. For this
state,
\begin{eqnarray}
\langle\colon T_{\alpha\beta}(x)\colon\rangle
&&=\langle \Psi|\colon T_{\alpha\beta}(x)\colon
|\Psi\rangle \nonumber \\
&&={\epsilon\over 1+\epsilon^2}\,\lbrace{\sqrt 2}
(T_{\alpha\beta}[f_{\bf k},
f_{\bf k}]+T_{\alpha\beta}[f_{\bf k}^{\ast},
f_{\bf k}^{\ast}])+2\epsilon(T_{\alpha\beta}[f_{\bf k},
f_{\bf k}^{\ast}]+T_{\alpha\beta}[f_{\bf k}
^{\ast},f_{\bf k}])\rbrace
\nonumber \\
&&={{{\cal K_{\alpha\beta}}\epsilon}\over{1+\epsilon^2}}
\big(2\epsilon-{\sqrt 2} \cos(2\theta)\big).  \label{eq:vacplus2rho}
\end{eqnarray}
Obviously the energy density can be positive or negative depending
on the value of $\epsilon$ and the
spacetime-dependent phase $\theta\equiv
k_{\rho}x^{\rho}$.
We also observe that the negative
contribution comes from the cross term. For a
general state which is a linear combination
of $N$ particle number eigenstates, the number
of cross terms will increase as $N^2$.
Therefore, for a general quantum state, the
occurrence of a negative energy density is very
probable. However, the total energy is always non-negative.
We will come back to this
simple illustrative example when discussing
the breakdown of the semiclassical gravity
theory later.

\section{Limits of the Semiclassical Theory}
\label{sec:limits}

To obtain a criterion of the validity of
semiclassical gravity theory, we may recall
the investigation in Ref. \cite{Ford82}. There the energy flux of
gravitational radiation in  linearized gravity
produced by a matter field was calculated both in the semiclassical
theory and in a linear quantum gravity theory. In the semiclassical
theory based upon
Eq.(\ref{eq:semi}), the flux depends on products of
expectation values of stress
tensor operators, whereas in a theory in which the metric perturbations are
quantized, it depends upon the corresponding products of expectation
values.
For the particular case of a single-mode coherent state, the product of
expectation values is the same as the expectation value of products of
stress tensors. This can be understood from the fact that a coherent state
corresponds to a classical field excitation, for which one would expect
the semiclassical theory to be a good approximation.

We propose that the extent to which the semiclassical approximation
is violated can be measured by
the dimensionless quantity
\begin{equation}
\Delta_{\alpha\beta\mu\nu}(x,y)
\equiv \Biggl|
{\langle\colon
T_{\alpha\beta}(x)\,T_{\mu\nu}(y)\colon\rangle
-\langle\colon T_{\alpha \beta}(x)\colon\rangle
\langle\colon T_{\mu \nu}(y)\colon\rangle
\over \langle\colon T_{\alpha\beta}(x)\,
T_{\mu\nu}(y)\colon\rangle }  \Biggr| .
\end{equation}
This quantity is a dimensionless measure of the stress tensor fluctuations.
(Note that it is not a tensor, but rather the ratio of tensor components.)
If its components are always small compared to unity, then these fluctuations
are small and we expect the semiclassical theory to hold.
However, the numerous components and the dependence upon two spacetime points
make this a rather cumbersome object to study. For simplicity, we will
concentrate upon the coincidence limit, $x \rightarrow y$, of the purely
temporal component of the above quantity, that is
\begin{equation}
\Delta(x)
\equiv \Biggl|
{
\langle\colon T_{00}{}^2(x)\colon
\rangle-\langle\colon T_{00}(x)\colon\rangle^2
\over
\langle\colon T_{00}{}^2(x)\colon\rangle}
\Biggr|.    \label{eq:Delta}
\end{equation}
The local energy density fluctuations are small when $\Delta \ll1$,
which we take to be a measure of the validity of the semiclassical theory.
Note that we have used normal ordering with respect to the Minkowski vacuum
state to define the various operators.

It is not difficult to see why
the semiclassical gravity theory is not expected to be
valid when the energy fluctuations are
large. Suppose we have a quantum state which is
a superposition of two states, each of which
describes a distinct macroscopic matter
configuration, e.g., (1) the presence of a 1000
kg mass, or (2) the absence of this mass.
Clearly, a measurement of the gravitational field
should reveal either (1) the field of a
1000 kg mass, or (2) no field, each with 50\%
probability. But the semiclassical field
equations predict finding the field of a 500
kg mass with 100\% probability.

Now we will discuss some specific examples
involving the massless scalar field in flat spacetime. First we consider
the general quantum state $|\Psi\rangle
=\sum_{n=0}^{\infty} c_n|n\rangle$, for which a single mode is excited. The
correlation function for the stress tensor can be divided into three parts,
\begin{equation}
\langle\colon
T_{\alpha\beta}\,T_{\mu\nu}\colon\rangle
=\langle\Psi|\colon T_{\alpha\beta}\,
T_{\mu\nu}\colon |\Psi\rangle = P_0 +P_2 +P_4.
\end{equation}
The contribution of the diagonal terms is
\begin{eqnarray}
&&P_0 = \sum_{n=0}^\infty |c_n|^2\,
(n-1)^{1/2}\,n\,(n+1)^{1/2} \nonumber \\
&&(T_{\alpha\beta}[f_{\bf k}^{\ast},
f_{\bf k}]\,T_{\mu\nu}[f_{\bf k}^{\ast},
f_{\bf k}]+ T_{\alpha\beta}[f_{\bf k}^{\ast},
f_{\bf k}]\,T_{\mu\nu}[f_{\bf k},
f_{\bf k}^{\ast}] \nonumber \\
&&+ T_{\alpha\beta}[f_{\bf k},
f_{\bf k}]\,T_{\mu\nu}[f_{\bf k}^{\ast},
f_{\bf k}^{\ast}]+ T_{\alpha\beta}[f_{\bf k},
f_{\bf k}^{\ast}]\,T_{\mu\nu}[f_{\bf k},
f_{\bf k}^{\ast}] \nonumber \\
&&+ T_{\alpha\beta}[f_{\bf k},
f_{\bf k}^{\ast}]\,T_{\mu\nu}[f_{\bf k}^{\ast},
f_{\bf k}]+T_{\alpha\beta}[f_{\bf k}^{\ast},
f_{\bf k}^{\ast}]\,T_{\mu\nu}[f_{\bf k},
f_{\bf k}]).
\end{eqnarray}
That of the matrix elements between $|n\rangle$ and
$|n+2\rangle$ is
\begin{eqnarray}
&&P_2 = \sum_{n=0}^\infty c_n\,c_{n+2}^{\ast}\,
n\,(n+1)^{1/2}\,(n+2)^{1/2} \nonumber \\
&&(T_{\alpha\beta}[f_{\bf
k},f_{\bf k}^{\ast}]\, T_{\mu\nu}[f_{\bf
k}^{\ast},f_{\bf k}^{\ast}]+
T_{\alpha\beta}[f_{\bf k}^{\ast},f_{\bf k}]\,
T_{\mu\nu}[f_{\bf k}^{\ast},f_{\bf
k}^{\ast}] \nonumber \\
&&+ T_{\alpha\beta}[f_{\bf k}^{\ast},f_{\bf
k}^{\ast}]\, T_{\mu\nu}[f_{\bf k},f_{\bf
k}^{\ast}] + T_{\alpha\beta}[f_{\bf k}^{\ast},
f_{\bf k}^{\ast}]\, T_{\mu\nu}[f_{\bf
k}^{\ast},f_{\bf k}]) + c.c.\, ,
\end{eqnarray}
whereas that between $|n\rangle$ and $|n+4\rangle$ is
\begin{equation}
P_4 = \sum_{n=0}^\infty c_n\,c_{n+4}\,(n+4)^{1/2}
(n+3)^{1/2}\,(n+2)^{1/2}\,(n+1)^{1/2}\,
T_{\alpha\beta}[f_{\bf k}^{\ast},f_{\bf
k}^{\ast}]\, T_{\mu\nu}[f_{\bf
k}^{\ast},f_{\bf k}^{\ast}] + c.c.\, ,
\end{equation}
where $c.c.$ denotes the complex conjugate.
The effect of the normal ordering has been to remove the contribution of
the $n=0$ term from the diagonal part, $P_0$.

\subsection{The Vacuum + Two Particle State}

Now we go back to the simple case of a state which is a superposition
of the vacuum and a two particle eigenstate, Eq. (\ref{eq:vacplus2}).
In this case we find
\begin{eqnarray}
\langle\colon T_{\alpha\beta}(x)\,
T_{\mu\nu}(x)\colon\rangle =&&
{2\epsilon^2\over (1+\epsilon^2)^2}\,
(T_{\alpha\beta}[f_{\bf k}^{\ast},f_{\bf
k}^{\ast}]\, T_{\mu\nu}[f_{\bf
k},f_{\bf k}] +
T_{\alpha\beta}[f_{\bf k}^{\ast},f_{\bf
k}]\, T_{\mu\nu}[f_{\bf
k}^{\ast},f_{\bf k}] \nonumber \\
&&+T_{\alpha\beta}[f_{\bf k}^{\ast},f_{\bf
k}]\, T_{\mu\nu}[f_{\bf
k},f_{\bf k}^{\ast}]+
T_{\alpha\beta}[f_{\bf k},f_{\bf
k}^{\ast}]\, T_{\mu\nu}[f_{\bf
k}^{\ast},f_{\bf k}] \nonumber \\
&&+T_{\alpha\beta}[f_{\bf k},f_{\bf
k}^{\ast}]\, T_{\mu\nu}[f_{\bf
k},f_{\bf k}^{\ast}]+
T_{\alpha\beta}[f_{\bf k},f_{\bf
k}]\, T_{\mu\nu}[f_{\bf
k}^{\ast},f_{\bf k}^{\ast}]) \nonumber \\
=&&{\cal K}_{\alpha\beta}\,{\cal K}_{\mu\nu}\:
{{12\epsilon^2}\over (1+\epsilon^2)^2}.
\end{eqnarray}
{}From this result, Eq. (\ref{eq:vacplus2rho}), and the
fact that ${\cal K}_{00}= {\omega\over 2L^3}$, we find that
\begin{equation}
\Delta(x)={{10\epsilon +\sqrt{2} cos(2\theta)}
\over {12\epsilon}}.
\end{equation}
{}From Eq.(\ref{eq:vacplus2rho}), it follows that
the condition for the expectation value
of energy density to be negative is
$cos(2\theta) > \sqrt{2}\epsilon$. In this case we have $\Delta(x)>1$.
This indicates when the energy density is negative, the energy density
fluctuations are large and the semiclassical theory is not a
good approximation.

\subsection{Squeezed States }

Squeezed states of light have been extensively
investigated recently in quantum optics and have been experimentally
realized\cite{Wu86}. Squeezing reduces the quantum
uncertainty in one variable with a corresponding increase in that of
its conjugate variable. Squeezed states form a two parameter family
of quantum states, which include both coherent states and states
with negative energy density as different limits.

A general squeezed state for a single mode can
be expressed as \cite{Caves81}
\begin{equation}
|\alpha,\zeta\rangle=D(\alpha)\,S(\zeta)
\,|0\rangle,
\end{equation}
where $D(\alpha)$ is the displacement operator
\begin{equation}
D(\alpha)\equiv exp(\alpha a^{\dagger}-
\alpha^{\ast}a)=e^{-|\alpha|^2/2}\,
e^{\alpha a^\dagger}\,e^{-\alpha^\ast a}
\end{equation}
and $S(\zeta)$ is the squeeze operator
\begin{equation}
S(\zeta)\equiv exp[{1\over 2}\zeta^\ast a^2
-{1\over 2}\zeta ({a^\dagger})^2].
\end{equation}
Here
\begin{equation}
\alpha = s e^{i\gamma}
\end{equation}
 and
\begin{equation}
\zeta = re^{i\delta}.
\end{equation}
are arbitrary complex numbers.
The displacement and squeeze operators satisfy the relations
\begin{equation}
D^{\dagger}(\alpha)\,a\,D(\alpha)=a+\alpha,
\end{equation}
\begin{equation}
D^{\dagger}(\alpha)\,a^{\dagger}\,D(\alpha)
=a^{\dagger}+\alpha^{\ast},
\end{equation}
\begin{equation}
S^{\dagger}(\zeta)\,a\,S(\zeta)=
a\,\cosh r-a^{\dagger}e^{i\delta}\sinh r,
\end{equation}
and
\begin{equation}
S^{\dagger}(\zeta)\,a^{\dagger}\,S(\zeta)=
a^{\dagger}\,\cosh r-ae^{-i\delta}\sinh r.
\end{equation}
Using the above formulae, the expectation value of the
renormalized stress tensor is easily obtained:
\begin{eqnarray}
\langle \alpha,\zeta|\colon T_{\alpha\beta}(x) \colon|\alpha,\zeta\rangle
=&&\langle 0|\, S^{\dagger}(\zeta)\,
D^{\dagger}(\alpha)\,
(a^2 T_{\alpha\beta}[f_{\bf k},f_{\bf k}]
+a^{\dagger}a\,(T_{\alpha\beta}
[f_{\bf k},f_{\bf k}^{\ast}] \nonumber \\
&&+T_{\alpha\beta}[f_{\bf k}^{\ast},f_{\bf k}])
+(a^{\dagger})^2 T_{\alpha\beta}
[f_{\bf k}^{\ast},f_{\bf k}^{\ast}])
\,D(\alpha)\,S(\zeta)\,|0\rangle \nonumber \\
=&& 2{\cal K}_{\alpha\beta} \bigl\{
\sinh r \cosh r \cos(2 \theta +\delta) +\sinh^2 r \nonumber \\
&&+ s^2 [1 - \cos 2(\theta+ \gamma)] \bigr\}
    \label{eq:T00}
\end{eqnarray}
Simlarly, the expectation value of the squared
stress tensor is
\begin{eqnarray}
\langle\alpha,\zeta|\colon T_{\alpha\beta}(x)T_{\mu\nu}(x)\colon
|\alpha,\zeta\rangle &&
= 2{\cal K_{\alpha\beta}}\,{\cal K_{\mu\nu}}
  \biggl( s^4 \bigl[ \cos 4(\theta+\gamma)
                    -4 \cos 2(\theta+\gamma) +3 \bigr] \nonumber \\
      &&  +3s^2 \Bigl\{ 2\sinh r \cosh r \bigl[ 2\cos(2\theta +\delta+2\gamma)
        -\cos(4\theta +\delta+2\gamma) \bigr]   \nonumber \\
&& + 4 \sinh^2 r (\cos 2\gamma - \cos 2\theta)
               -\cos(\delta +2\gamma) \Bigr\}   \nonumber \\
  &&  +3 \sinh^2 r \bigl[ \cosh^2 r \cos (4\theta +2\delta) +
        3 -4\cos 2\theta \bigr] \biggr).     \label{eq:T00sq}
\end{eqnarray}

The degree of squeezing may be measured by the
squeeze parameter $|\zeta|=r$. When $\zeta=0$, the squeezed states reduce
to coherent states. In this particular case, the expectation value of
a product of stress tensors is equal to the corresponding product of
expectation values \cite{Ford82}:
\begin{equation}
\langle\colon T_{\alpha \beta}(x) \,T_{\mu \nu}(y)\colon\rangle =
\langle\colon T_{\alpha \beta}(x)\colon\rangle
\langle\colon T_{\mu \nu}(y)\colon\rangle,
\end{equation}
and hence
\begin{equation}
\Delta_{\alpha\beta\mu\nu}(x,y) = \Delta = 0.
\end{equation}
This result is consistent with the interpretation of coherent states as the
quantum states which describe classical field excitations. By the criterion
which we have adopted, the semiclassical gravity theory is a good
approximation for coherent states.

The opposite limit from a coherent state is the case where $\alpha =0$,
known as a squeezed vacuum state. Such a state is not, of course, the vacuum
state so long as $\zeta \not= 0$, but rather a superposition of states
containing even numbers of particles. Squeezed vacuum states have a
particular physical interest because they are the states which result
from quantum particle creation processes. Such states of the electromagnetic
field have been generated in the laboratory using nonlinear optical media,
and have been the topic of much interest in quantum optics in recent years.

In a squeezed vacuum, $\alpha=0$, we may also take $\delta=0$, as this
is simply a choice of phase, and write
\begin{equation}
\langle \alpha,\zeta|\colon
T_{\alpha\beta}(x) \colon|\alpha,\zeta\rangle
=2{\cal K}_{\alpha\beta} \sinh r [\cosh r
\cos(2\theta)+ \sinh r].
\end{equation}
Here $r=0$ corresponds to the vacuum state and hence
gives a vanishing expectation value for the stress tensor. For $r \not =0$,
the squeezed vacuum state exhibits negative energy densities. That is, for
fixed $r$ as $\theta$ varies from $0$ to $2\pi$ (either the spatial position
changing by one wavelength at fixed time, or the time varying through one
period at fixed position), the energy density becomes negative during
part of the cycle.

For the case $\delta=0$ and $\alpha=0$, the expectation value of the
squared stress tensor is
\begin{eqnarray}
\langle\alpha,\zeta|\colon
T_{\alpha\beta}(x)T_{\mu\nu}(x)\colon
|\alpha,\zeta\rangle
=&& 2{\cal K_{\alpha\beta}}\,{\cal K_{\mu\nu}}
\sinh^2 r \lbrack 2 \cosh^2 r  \cos 4\theta
-8 \sinh r \cosh r \cos 2\theta \nonumber \\
&&+3(\sinh^2 r+ \cosh^2 r)
\rbrack.
\end{eqnarray}
As required, for the vacuum state ($r=0$) this
quantity vanishes.

{}From the above results, we can form the quantity $\Delta$. However, the
analytical expressions are not particularly transparent. Our primary
concern is whether or not $\Delta \ll 1$, which is best determined by
numerical evaluation of $\Delta$ using Eqs.~(\ref{eq:Delta}),
(\ref{eq:T00}), and (\ref{eq:T00sq}). The figures illustrate the
results. For states which are
sufficiently close to coherent states, i.e. $r \ll |\alpha|$,
we do indeed find that $\Delta \ll 1$. However, as the magnitude
of the squeeze parameter $r$ increases relative to that of $|\alpha|$,
we find that $\Delta$ increases. By the point that the state is
sufficiently squeezed to have
$\rho = \hbox{$\langle :T_{00}: \rangle$} < 0$, we
always have that $\Delta$ is at least of order unity. Thus squeezed states
for which the energy density is negative exhibit large energy density
fluctuations.

\subsection{The Casimir Vacuum}

One of the most astonishing predictions of
quantum field theory is the Casimir effect
\cite{Casimir48}, in which the vacuum energy density
of the quantized electromagnetic field between two parallel
perfectly  conducting plates is negative. We will
consider the Casimir effect for a massless, minimally coupled
scalar field. The expectation value of the stress tensor can be
expressed as
\begin{equation}
\langle T_{\mu\nu}(x)\rangle =
{1\over 2}\lim_{x'\to x}
(\nabla_{\mu'} \nabla_\nu+
\nabla_\mu \nabla_{\nu'}-
g_{\mu\nu}\nabla_\alpha\nabla^{\alpha'})
\,G(x,x'),
\end{equation}
where the Hadamard elementary function
$G(x,x')$ is
\begin{equation}
G(x,x')={1\over 2}\langle\phi(x)\phi(x')
+\phi(x')\phi(x)\rangle.
\end{equation}
This quantity is formally infinite and is renormalized by replacing
$G(x,x')$ by the renormalized Green's function
\begin{equation}
G_R(x,x') = G(x,x') - G_0(x,x'),
\end{equation}
where $G_0(x,x')$ is the Minkowski space Green's function, i.e., that
in the absence of boundaries. The resulting finite stress energy is
the expectation value in the Casimir vacuum of $T_{\mu\nu}$ normal
ordered with respect to the Minkowski vacuum:
\begin{equation}
\langle :T_{\mu\nu}(x):\rangle_C =
{1\over 2}\lim_{x'\to x}
(\nabla_{\mu'} \nabla_\nu+
\nabla_\mu \nabla_{\nu'}-
g_{\mu\nu}\nabla_\alpha\nabla^{\alpha'})
\,G_R(x,x').
\end{equation}

For the purpose of illustration, we will consider the particular
boundary condition of periodicity in the $z$-direction with periodicity
length $L$. In this case, the Green's function can be expressed
as an image sum:
\begin{equation}
G(x,x')=\sum_{n=-\infty}^\infty
G_n(\sigma_n(x,x')),   \label{eq:GF}
\end{equation}
where
\begin{equation}
G_n(\sigma_n(x,x'))=
{1\over{4\pi^2\sigma_n}}.
\end{equation}
The geodesic distance $\sigma$ for different
$n$ is
\begin{equation}
\sigma_n(x,x')=
-(t-t')^2+({\bf r}-{\bf r}_n')^2,
\end{equation}
where
\begin{equation}
x=(t,{\bf r}), \quad
x'=(t', {\bf r}'), \quad
{\bf r}_n'=(x',y',z'-nL).
\end{equation}
The renormalized Green's function $G_R(x,x')$ is simply given by
Eq. \ (\ref{eq:GF}) with the $n=0$ term omitted.

Thus we find
\begin{eqnarray}
\langle\colon T_{\mu\nu}(x)\colon\rangle_C
&&={1\over{8\pi^2}}\lim_{x'\to x}
\sum_{n=-\infty}^\infty
(\partial_{\mu'} \partial_\nu+
\partial_{\mu} \partial_{\nu'}+
\eta_{\mu\nu}\eta^{\rho\sigma}
\partial_\rho \partial_{\sigma'})
\sigma_n^{-1} \nonumber \\
&&=-{\zeta(4)\over{\pi^2 L^4}}\:
diag[1,-1,-1,3], \label{eq:stresstensor}
\end{eqnarray}
where $\zeta(s)=\sum_{n=1}^{\infty}
{1\over{n^s}}$ is the
Riemann zeta function, and we have excluded
the $n=0$ term, since it is just the
contribution from the Minkowski vacuum.
Since $\zeta(4) ={\pi^4\over 90}$,
\begin{equation}
\langle\colon T_{00}(x)\colon\rangle
=-{\pi^2\over{90\,L^4}}\,.
\end{equation}

    We now proceed to define the stress tensor correlation
function as before by interpreting $\colon
T_{\alpha\beta}(x)\,T_{\mu\nu}(y)\colon$ as being
normal ordered with respect to the Minkowski vacuum state.
Its expectation value is obtained by differentiation of the
corresponding four-point function
\begin{eqnarray}
\langle\colon T_{\alpha\beta}(x)
T_{\mu\nu}(y)\colon\rangle=&&
\lim_{x_1,x_2 \to x \atop x_3,x_4\to y}\,
(\eta_{\alpha\lambda}\eta_{\beta\gamma}-
{1\over 2}\eta_{\alpha\beta}
\eta_{\lambda\gamma})\,
(\eta_{\mu\rho}\eta_{\nu\sigma}-
{1\over 2}\eta_{\mu\nu}
\eta_{\rho\sigma}) \times \nonumber \\
&&\partial_1^\lambda\,
\partial_2^\gamma\,
\partial_3^\rho\,
\partial_4^\sigma\,
\langle :\phi(x_1)\phi(x_2)\phi(x_3)\phi(x_4):
\rangle_C.
\end{eqnarray}
This four-point function is the expectation value of an operator
normal ordered with respect to the Minkowski vacuum, but whose
expectation value is taken in the Casimir vacuum state. We have
in fact dealt with similar quantities in our calculation of the
mean stress tensor, namely $\langle\colon T_{\mu\nu}\colon\rangle_C$
itself and the renormalized Green's function
\begin{equation}
G_R(x,x')={1\over 2}\langle :\phi(x)\phi(x')
+\phi(x')\phi(x): \rangle_C.
\end{equation}

The evaluation of the expectation value of the four-point function
can be achieved with the aid of Wick's theorem. Let $\phi_i = \phi(x_i)$.
The quartic product may be expressed as
\begin{eqnarray}
\phi_1\phi_2\phi_3\phi_4 = &&:\phi_1\phi_2\phi_3\phi_4: +
:\phi_1\phi_2: \langle\phi_3\phi_4\rangle_M +
:\phi_1\phi_3: \langle\phi_2\phi_4\rangle_M +
:\phi_1\phi_4: \langle\phi_2\phi_3\rangle_M + \nonumber \\
&&:\phi_2\phi_3: \langle\phi_1\phi_4\rangle_M +
:\phi_2\phi_4: \langle\phi_1\phi_3\rangle_M +
:\phi_3\phi_4: \langle\phi_1\phi_2\rangle_M +  \nonumber \\
&&\langle\phi_1\phi_2\rangle_M \langle\phi_3\phi_4\rangle_M +
\langle\phi_1\phi_3\rangle_M \langle\phi_2\phi_4\rangle_M +
\langle\phi_1\phi_4\rangle_M \langle\phi_2\phi_3\rangle_M,
\end{eqnarray}
where $\langle \rangle_M$ is the expectation value in the
Minkowski  vacuum. However, it may equally well be written as
\begin{eqnarray}
\phi_1\phi_2\phi_3\phi_4 = &&N_C(\phi_1\phi_2\phi_3\phi_4) +
N_C(\phi_1\phi_2) \langle\phi_3\phi_4\rangle_C +
N_C(\phi_1\phi_3) \langle\phi_2\phi_4\rangle_C + \nonumber \\
&&N_C(\phi_1\phi_4) \langle\phi_2\phi_3\rangle_C +
N_C(\phi_2\phi_3) \langle\phi_1\phi_4\rangle_C + \nonumber \\
&&N_C(\phi_2\phi_4) \langle\phi_1\phi_3\rangle_C +
N_C(\phi_3\phi_4) \langle\phi_1\phi_2\rangle_C +  \nonumber \\
&&\langle\phi_1\phi_2\rangle_C \langle\phi_3\phi_4\rangle_C +
\langle\phi_1\phi_3\rangle_C \langle\phi_2\phi_4\rangle_C +
\langle\phi_1\phi_4\rangle_C \langle\phi_2\phi_3\rangle_C,
\end{eqnarray}
where $N_C$ denotes normal ordering with respect to the Casimir vacuum.
Recall that Wick's theorem applies to any quantum state $|\psi\rangle$
for which there is a decomposition of the field operator into positive
and negative frequency parts, $\phi = \phi^{+} + \phi^{-}$, so that
$\phi^{+} |\psi\rangle = \langle\psi| \phi^{-} =0$. Both the Minkowski
and Casimir vacua satisfy this condition. We now wish to equate the
two expressions above, solve for $:\phi_1\phi_2\phi_3\phi_4:$,
and take its expectation value in the Casimir vacuum. If we use such
relations as $\langle N_C(\phi_1\phi_2)\rangle_C =0$ and
\begin{equation}
\langle \phi_1\phi_2\rangle_M = \langle \phi_1\phi_2\rangle_C
 - \langle N_M(\phi_1\phi_2) \rangle_C,
\end{equation}
we obtain the result
\begin{eqnarray}
\langle :\phi(x_1)\phi(x_2)\phi(x_3)\phi(x_4):\rangle_C =
\langle\colon\phi(x_1)\phi(x_2)\colon\rangle_C\,
\langle\colon\phi(x_3)\phi(x_4)\colon\rangle_C+
\nonumber \\
\langle\colon\phi(x_1)\phi(x_3)\colon\rangle_C\,
\langle\colon\phi(x_2)\phi(x_4)\colon\rangle_C+
\langle\colon\phi(x_1)\phi(x_4)\colon\rangle_C\,
\langle\colon\phi(x_2)\phi(x_3)\colon\rangle_C.
\end{eqnarray}

Using this result, we may express the expectation value of the squared
energy density as
\begin{eqnarray}
\langle\colon T_{00}{}^2(x)\colon\rangle_C =&&
\lim_{x_1,x_2,x_3,x_4\to x}\,
(\delta_{0\lambda}\,\delta_{0\gamma} -
{1\over 2}\eta_{\lambda\gamma})\,
(\delta_{0\rho}\,\delta_{0\sigma} -
{1\over 2}\eta_{\rho\sigma})\,
\partial_1^\lambda\,
\partial_2^\gamma\,
\partial_3^\rho\,
\partial_4^\sigma\,\times \nonumber \\
&&\Bigl(\langle\colon\phi(x_1)\phi(x_2)\colon\rangle\,
\langle\colon\phi(x_3)\phi(x_4)\colon\rangle+
\langle\colon\phi(x_1)\phi(x_3)\colon\rangle\,
\langle\colon\phi(x_2)\phi(x_4)\colon\rangle+ \nonumber \\
&&\langle\colon\phi(x_1)\phi(x_4)\colon\rangle\,
\langle\colon\phi(x_2)\phi(x_3)\colon\rangle \Bigr). \nonumber \\
=&& \langle\colon T_{00}\colon\rangle_C^2 +
 {1 \over 2}\bigl[
 \langle\colon {\dot \phi}^2 \colon\rangle_C^2 +
\langle\colon (\phi_{,x})^2 \colon\rangle_C^2 +
\nonumber \\
&&\langle\colon (\phi_{,y})^2 \colon\rangle_C^2 +
\langle\colon (\phi_{,z})^2 \colon\rangle_C^2 \bigr].
\end{eqnarray}
Let $\rho =\langle\colon T_{00}\colon\rangle_C$ be the Casimir
energy density and let $p_i =\langle\colon T_{ii}\colon\rangle_C$
be the pressure in the $i$-direction. Then for the massless scalar
field we have
\begin{equation}
 \langle\colon {\dot \phi}^2 \colon\rangle_C^2 =
{1 \over 2} (\rho +p_1 +p_2 +p_3)^2
\end{equation}
and, for example,
\begin{equation}
 \langle\colon (\phi_{,x})^2 \colon\rangle_C^2 =
{1 \over 2} (\rho +p_1 -p_2 -p_3)^2.
\end{equation}
{}From these relations, we may write
\begin{eqnarray}
\Delta' \equiv &&{{\langle\colon T_{00}{}^2(x)\colon\rangle -\rho^2}
\over \rho^2} = {1\over 8}
\bigl[ (\rho +\xi_1 +\xi_2 +\xi_3)^2 + \nonumber \\
&&(\rho +\xi_1 -\xi_2 -\xi_3)^2 + (\rho -\xi_1 +\xi_2 -\xi_3)^2
+ (\rho -\xi_1 -\xi_2 +\xi_3)^2 \bigl], \label{eq:deltaprime}
\end{eqnarray}
where $\xi_i = p_i/\rho$.
The quantity $\Delta'$ is related to $\Delta$ by
\begin{equation}
\Delta' = {\Delta \over {1 -\Delta}}, \qquad
\Delta = {\Delta' \over {1 +\Delta'}},
\end{equation}
and is an equivalent measure of the scale of the energy density
fluctuations.

We may obtain  lower bounds on $\Delta'$ and $\Delta$. The minimum value
of the right hand side of Eq.\ (\ref{eq:deltaprime}) is $1/2$ and occurs
when $\xi_1=\xi_2=\xi_3 =0$. Thus we have that
\begin{equation}
\Delta' \ge {1 \over 2}, \qquad  \Delta \ge {1 \over 3}.
\end{equation}
This means that the dimensionless Casimir energy density fluctuations
are always at least of order unity. Note that this bound is independent
of the details of the geometry of the boundaries. For the particular
case of periodicity in one spatial direction, we have from
Eq.\ (\ref{eq:stresstensor}) that $\xi_1=\xi_2 =-1$ and $\xi_3=3$, so
that $\Delta' =6$ and $\Delta = 6/7$. The essential point here is that
neither measure of the fluctuations is small, and hence we conclude
that the gravitational field due to Casimir energy is not described
by a fixed classical metric. That the Casimir force is a
fluctuating force, and that Casimir's calculation yields the mean value
of that force, have been emphasized by Barton \cite{Barton91}.

\section{Metric Fluctuations and Test Particles}
\label{sec:test}

The properties of spacetime can be probed
by  test particles which follow geodesics
in a classical gravitational field. When the
gravitational field is described by a classical metric,
we can use test particles as an operational probe of the
geometry. Our problem is now to give a meaning to the gravitational
field of a fluctuating source, such as the Casimir vacuum. We
propose to replace the description in which the test particles follow
fixed trajectories by a statistical description in which one only
attempts to compute average quantities such as the mean squared
velocity of an ensemble of test particles. This is the approach which
is used to describe Brownian motion by means of a Langevin equation.
The basic idea is that the particle is subjected both to a classical
force and to a fluctuating force. For nonrelativistic motion, its
equation of motion may be written as
\begin{equation}
m{d{\bf v}(x)\over dt}= {\bf F}_c(x) + {\bf F}(x),
\end{equation}
where $m$ is the test particle mass, ${\bf F}_c$ is the classical force,
and ${\bf F}$ is the fluctuating force.
The solution of this equation is
\begin{equation}
{\bf v}(t)= {\bf v}(t_0) + {1\over m}\int_{t_0}^t
  [{\bf F}_c(t') + {\bf F}(t')] \,dt'.
\end{equation}
We assume that the fluctuating force averages to zero,
$\langle {\bf F}\rangle =0$, but that quantities quadratic in ${\bf F}$
do not. Thus the mean squared velocity, averaged over an ensemble of
test particles is, for the case that ${\bf v}(t_0) =0$,
\begin{equation}
\langle {\bf v}^2\rangle = {1\over m^2}
\int_{t_0}^t dt_1 \int_{t_0}^t dt_2\,
[{\bf F}_c(t_1)\,{\bf F_c}(t_2) +
\langle {\bf F}(t_1)\,{\bf F}(t_2)\rangle]. \label{eq:vsquare}
\end{equation}
Typically, the correlation function for the fluctuating force vanishes
for times separated by much more than some correlation time, $t_c$.
In this case, the contribution of the fluctuating force to
$\langle {\bf v}^2\rangle$ grows linearly in time.

We wish to consider the motion of a test particle in a weak
gravitational field, so
\begin{equation}
g_{\mu\nu}=\eta_{\mu\nu}+h_{\mu\nu},
\end{equation}
where $|h_{\mu\nu}|\ll 1$.
We further assume that the particle's motion is non-relativistic.
However, we make no assumptions concerning the relative magnitudes
of the stress tensor components, in contrast to the usual Newtonian
limit where one assumes that $T_{00}$ is large compared to
all other components.
The proper time interval along the particle's world line is
\begin{eqnarray}
d\tau^2&&=g_{\mu\nu}\,dx^\mu\,dx^\nu
=g_{00}\,dt^2 + g_{ij}\,dx^i\,dx^j \nonumber \\
&&=dt^2\,(g_{00}+2g_{0i}\,v^i+
g_{ij}\,v^i\,v^j),
\end{eqnarray}
where $v^i\equiv dx^i/dt$ is the velocity of
the test particle.
The dynamics of the test particle is
determined by the variational principle
\begin{equation}
\delta\int
\bigl({d\tau\over dt}\bigr)\,dt=0.
\end{equation}
Therefore, the Lagrangian of the test particle
is
\begin{eqnarray}
L&&={d\tau\over dt} \nonumber \\
&&=(-1+h_{00}+2h_{0i}v^i+{\bf v}^2
+h_{ij}\,v^i\,v^j)^{1/2}.
\end{eqnarray}
Up to terms linear in $h_{\mu\nu}$, the
Lagrangian is
\begin{equation}
L=-1+{1\over 2}h_{00}+h_{0i}\,v^i+
{1\over 2}{\bf v}^2
+{1\over 2}h_{ij}\,v^i\,v^j.
\end{equation}
The curvature tensor is related to
$h_{\mu\nu}$ through
\begin{equation}
R_{\mu\nu}={1\over 2}
(h_{\!\nu\alpha}{}{}^{,\alpha}{}_{,\mu}
+h_{\!\mu\alpha}{}{}^{,\alpha}{}_{,\nu}
-\partial_\rho\partial^\rho\,h_{\mu\nu}
-h_{,\mu\nu}),
\end{equation}
where $h=h^\alpha {}_\alpha$.
In the Lorentz-Hilbert gauge,
\begin{equation}
\bar h_{\mu\nu}{}{}^{,\nu}\equiv
(h_{\mu\nu}-{1\over 2}\eta_{\mu\nu}\,h
^\alpha{}_\alpha)^{,\nu}=0,
\end{equation}
$h_{\mu\nu}$ can be determined by the
gravitational source through the
gravitational field equations
\begin{equation}
\partial_\rho\partial^\rho \,h_{\mu\nu}
=-16\pi G_N \bar T_{\mu\nu}
\equiv -16\pi G_N\,(T_{\mu\nu}-{1\over 2}
\eta_{\mu\nu}T^\alpha {}_\alpha).
\end{equation}
If we separate the spatial and temporal
components, we have
\begin{equation}
\partial_\rho\partial^\rho \,h_{00}
=-8\pi G_N\,(T_{00}+T^i{}_i),
\end{equation}
\begin{equation}
\partial_\rho\partial^\rho \,h_{0i}
=-16\pi G_N\,T_{0i},
\end{equation}
and
\begin{equation}
\partial_\rho\partial^\rho \,h_{ij}
=-16\pi G_N\,(T_{ij}+{1\over 2}\delta_{ij}T_{00}
-{1\over 2}\delta_{ij}T^k{}_k).
\end{equation}
Define generalized potentials, $\Phi$, $\Psi$, $\zeta_i$, and $\xi_{ij}$
as linear combinations of the metric perturbations so that
\begin{equation}
h_{00} = 2\Phi+2\Psi,
\end{equation}
\begin{equation}
h_{0i} = \zeta_i,
\end{equation}
\begin{equation}
h_{ij} =
\xi_{ij}-2\Phi\delta_{ij}+2\Psi\delta_{ij}.
\end{equation}
The potentials satisfy
\begin{mathletters}
\begin{equation}
\partial_\rho\partial^\rho\,\Phi = -4\pi G_N T_{00},
  \label{eq:Phi}
\end{equation}
\begin{equation}
\partial_\rho\partial^\rho\,\Psi = -4\pi G_N T^i{}_i,
   \label{eq:Psi}
\end{equation}
\begin{equation}
\partial_\rho\partial^\rho\,\zeta_i = -16\pi G_N T_{0i},
   \label{eq:zeta}
\end{equation}
and
\begin{equation}
\partial_\rho\partial^\rho\,\xi_{ij} = -16\pi G_N T_{ij}.
    \label{eq:xi}
\end{equation}
\end{mathletters}

The Lagrangian of the test particle is
\begin{equation}
L= -1 +(\Phi+\Psi)+\zeta_i\,v^i+
{1\over 2}{\bf v}^2\,(1-2\Phi+2\Psi)
+{1\over 2}\xi_{ij}\,v^i\,v^j.
\end{equation}
The Euler-Lagrange equation derived from this
Lagrangian, after discarding terms higher than
linear in the velocity or its first derivatives, is
\begin{equation}
\partial_i\Phi+\partial_i\Psi
+\partial_i(\zeta_i\,v^j)-\dot\zeta_i
+\dot v_i(1-2\Phi+2\Psi)+
v_i\,(-2\dot\Phi+2\dot\Psi)+\dot\xi_{ij}\,v^j
+\xi_{ij}\,\dot v^j=0.
\end{equation}

We can now solve for the time derivative of velocity and
notice that in the weak field limit
\begin{equation}
(1-2\Phi+2\Psi+\xi_{ij})^{-1}\approx
1+2\Phi-2\Psi-\xi_{ij},
\end{equation}
so
\begin{equation}
\dot v_i=\dot\zeta_i-\partial_i\Phi
-\partial_i\Psi+
(2\dot\Phi\,\delta_{ij}-2\dot\Psi\,\delta_{ij}
-\partial_i\zeta_j-\dot\xi_{ij})\,v_j
-\zeta_j(\partial_i\,v_j).
\end{equation}
To lowest order
\begin{equation}
v_i=\zeta_i-\int(\partial_i\Phi(x))\,dt
-\int(\partial_i\Psi(x))\,dt. \label{eq:velocity}
\end{equation}

Equations (\ref{eq:Phi}- \ref{eq:xi}) may be solved in terms
of the retarded Green's function,
\begin{equation}
G_{ret}(x)=-{1\over 4\pi |{\bf x}|}\,
\,\delta(|{\bf x}-t|),
\end{equation}
to yield
\begin{mathletters}
\begin{equation}
\Phi(t, {\bf x})=G_N\int d^3x'\,
{T_{00}(t-|{\bf x}-{\bf x'}|,{\bf x'})\over
|{\bf x}-{\bf x'}|},
\end{equation}
\begin{equation}
\Psi(t, {\bf x})=G_N\int d^3x'\,
{T^i{}_i(t-|{\bf x}-{\bf x'}|,{\bf x'})\over
|{\bf x}-{\bf x'}|},
\end{equation}
\begin{equation}
\zeta_i(t, {\bf x})=4G_N\int d^3x'\,
{T_{0i}(t-|{\bf x}-{\bf x'}|,{\bf x'})\over
|{\bf x}-{\bf x'}|},
\end{equation}
and
\begin{equation}
\xi_{ij}(t, {\bf x})=4G_N\int d^3x'\,
{T_{ij}(t-|{\bf x}-{\bf x'}|,{\bf x'})\over
|{\bf x}-{\bf x'}|}.
\end{equation}
\end{mathletters}
The expression for the velocity, Eq. (\ref{eq:velocity}), now becomes
\begin{eqnarray}
v_i(t)&& = 4G_N\int d^3x'\,
{T_{0i}(t-|{\bf x}-{\bf x'}|,{\bf x'})\over
|{\bf x}-{\bf x'}|}
\nonumber \\
&&-G_N\,\partial_i \int d^3x'\int^t dt'\,
{T_{00}(t'-|{\bf x}-{\bf x'}|,{\bf x'})+
T^i{}_i(t'-|{\bf x}-{\bf x'}|,{\bf x'})\over
|{\bf x}-{\bf x'}|}.
\end{eqnarray}
The mean square of the $i$-component of the velocity is then
\begin{eqnarray}
\langle v_i{}^2(t) \rangle= && 16 C_{0i0i}(t,t)
        + \partial_i^2 \int^t dt' dt'' C_{0000}(t',t'') \nonumber \\
     &&  + \partial_i^2 \int^t dt' dt'' {{{C_j}^j}_k}^k (t',t'')
        -8 \partial_i \int^t dt' C_{0i00}(t,t') \nonumber \\
     && -8 \partial_i \int^t dt' {C_{0ij}}^j(t,t')
        -8 \partial_i^2 \int^t dt' dt'' {C_{00j}}^j(t',t''),
           \label{eq:vsquare2}
\end{eqnarray}
where
\begin{equation}
C_{\mu\nu\alpha\beta}(t',t'') \equiv G_N^2 \int d^3x' d^3x''
{\langle\colon T_{\mu\nu}(t'-|{\bf x}-{\bf x'}|,
{\bf x'})\,T_{\alpha\beta}(t''-|{\bf x}-{\bf x''}|,{\bf x''})\colon
\rangle\over |{\bf x}-{\bf x'}|\,|{\bf x}-{\bf x''}|}.
\end{equation}
Note that in Eq. (\ref{eq:vsquare2}), the index $i$ is not summed,
but all other repeated indices are summed.

This expression is rather cumbersome to evaluate explicitly,
but we can make order of magnitude estimates. Note that it is
of the general form of Eq. (\ref{eq:vsquare}), where the correlation
function of the fluctuating force is of order
\begin{equation}
\langle F F \rangle \approx m^2 C,
\end{equation}
where $C$ is a typical component of $C_{ijkl}$.
If $t_c$ is the correlation time, then we obtain the estimate
\begin{equation}
\langle v_i{}^2(t) \rangle \approx C t_c t.
\end{equation}

To be specific, let us consider a test particle moving in the fluctuating
gravitational field produced by the Casimir vacuum. We assume that
the particle is initially shot parallel
to and midway between two plates. Classically, the particle
will continue to move midway between the plates, as the gravitational
force vanishes by symmetry. However, it is subjected to a fluctuating
force which tends to cause it to drift toward one side or the other.
Since the energy density is of the order of
$1/L^4$, the characteristic force on the test particle is of order
$G_N m /L^3$, so $C \approx G_N^2 /L^6$. Furthermore, the correlation
time is of order $L$, the only length scale in the problem. Therefore,
\begin{equation}
\langle {v_i}^2(t)\rangle \approx
\bigg({L_P\over L}\bigg)^4
{t \over L}\, ,
\end{equation}
where $L_P$ is the Planck length. Thus the transverse velocity of the
test particle undergoes a random walk, with the rms transverse velocity
growing as the square root of time. We have of course restricted our
attention to a particular initial condition for the test particles, but
a similar analysis could be performed for any other initial condition.
The resulting probabilistic statements about such quantities as
$\langle {v_i}^2(t)\rangle$ sum up all that we can know about a
fluctuating gravitational field, such as that whose source is the
Casimir vacuum energy.

\section{Discussion and Conclusions}
\label{sec:last}

In this paper, we have argued that the semiclassical theory of
gravitation based upon Eq.(\ref{eq:semi}) is an approximation which
assumes that the fluctuations in $T_{\mu\nu}$ are small. This
assumption can fail far away from the Planck scale. The approximation
is valid for systems described by coherent states, but breaks down
for more general quantum states. In particular, states of a quantized
field in which the mean energy density is negative seem to exhibit large
energy density fluctuations. This was illustrated in our model calculations
for squeezed states and for the Casimir vacuum state. Thus the gravitational
field of such a system is not described by a fixed classical metric, but
rather by a fluctuating metric.

Because the operational significance of a gravitational field lies in
its effects upon test particles, the meaning of a fluctuating gravitational
field is given by a statistical description of ensembles of test particles.
This notion was developed in Sec.~\ref{sec:test}, where we proposed
a Langevin-type equation to determine such quantities as the
mean squared velocity of the test particles.

The analysis in this paper was restricted to a flat spacetime background.
In this case, the expectation values of quartic operators which appear
in the dimensionless measure of the fluctuations, $\Delta$, or in the
Langevin equation, Eq.(\ref{eq:vsquare2}), may be defined in terms of
normal ordered products. The generalization of this analysis to curved
spacetime backgrounds will require a renormalization procedure for
quartic operator products in such spacetimes, which has not yet been
developed. Note that an alternative approach to fluctuations
is to discuss only averaged quantities. If we were to average fields
over finite space or time intervals, then we could define a quantity
similar to $\Delta$ without invoking normal ordering. This approach
is used by Barton \cite{Barton91} in his treatment of the fluctuations
of the Casimir force. In the present context, this procedure suffers from
the defect that it introduces an arbitrary length or time scale into
the problem. If, for example, one wished to obtain an equation of the
form of Eq.(\ref{eq:vsquare}) using averaged rather than normal-ordered
quantities, there would be an inherent ambiguity in the resulting
equation.

A topic of considerable current interest is the possibility of
causality violation by a Lorentzian wormhole. Morris, Thorne, and
Yurtsever \cite{Morris88} have shown how such a wormhole might in
principle be constructed using the Casimir vacuum energy. A violation
of the weak energy condition, and hence negative energy densities, is
essential for the existence of a wormhole. Hochberg and Kephart
\cite{Hochberg} have suggested squeezed states as a possible source
for the negative energy, but an explicit wormhole model using
squeezed states has not yet been constructed. All of the literature
on wormhole models assumes the semiclassical gravity theory. However,
in light of the considerations in the present paper, it seems to
be important to include the possible effects of metric fluctuations.
Whether they are sufficiently large to prevent the creation of a
wormhole or its use to violate causality is not clear. Another
question of interest is whether a breakdown of the semiclassical
theory provides an alternative explanation as to why negative energy
fluxes cannot lead to violations of cosmic censorship \cite{Roman90}.

\acknowledgements

The work was supported in part by the National Science Foundation
under Grant PHY-9208805.

\figure{The energy density $\rho$ is plotted as a function of the
squeeze parameter $r$ and of $s = |\alpha|$. Here $\gamma = \delta =0$
and $\theta = \pi/2$.  When $s >> r$, the state is
close to a coherent state, and $\rho > 0$. However, when $s << r$, the
energy density is negative. \label{Fig1}}

\figure{Here $\Delta$ is plotted for the same range and choices of
parameters as in Fig. ~\ref{Fig1}. Note that $\Delta << 1$ only
when  $s >> r$, the region of classical behavior. Otherwise,
$\Delta \approx 1$.  \label{Fig2}}

\figure{The energy density $\rho$ is plotted as a function of the
squeeze parameter $r$ and of $\gamma$, which is the phase of $\alpha$.
Here $\delta =0$, $s = {1 \over 2}$, and $\theta = \pi/2$.
   \label{Fig3}}

\figure{Here $\Delta$ is plotted for the same range and choices of
parameters as in Fig. ~\ref{Fig3}. Again $\Delta << 1$ only
when  $r << 1$.  \label{Fig4}}

\end{document}